\documentclass[prl,showpacs,footinbib,twocolumn,final]{revtex4}
\usepackage{amsmath}
\usepackage{amssymb}
\usepackage{epsfig}
\usepackage{enumerate}

\newcommand{\TK}{T_{\rm K}}
\newcommand{\TKS}{T_{\rm K}^S}
\newcommand{\RRm}{\rho_{\rm m}}
\newcommand{\muB}{\mu_{\rm B}}
\newcommand{\subi}{{\rm m}}
\newcommand{\gm}{\gamma_{\rm m}}
\newcommand{\gammaphi}{\gamma_\phi}

\usepackage{graphics,graphicx,color}
\usepackage{times}
\usepackage{graphics,graphicx,color}

\begin{document}
  \title{Kondo decoherence: finding the right spin model for iron
    impurities in gold and silver
  }

\author{
  T. A. Costi$^{1,2}$, L. Bergqvist$^{1}$,
  A. Weichselbaum$^{3}$, J. von Delft$^{3}$,
  T. Micklitz$^{4,7}$, A. Rosch$^{4}$,\\
  P. Mavropoulos$^{1,2}$, P. H. Dederichs$^{1}$, 
  F. Mallet$^{5}$, L.  Saminadayar$^{5,6}$, and C. B\"auerle$^{5}$}
\affiliation{$^{1}$Institut f\"{u}r
      Festk\"{o}rperforschung, Forschungszentrum J\"{u}lich, 52425
      J\"{u}lich, Germany}
\affiliation{$^{2}$Institute for Advanced Simulation,
      Forschungszentrum J\"{u}lich, 52425
      J\"{u}lich, Germany}
\affiliation{$^{3}$Physics Department, Arnold
      Sommerfeld Center for Theoretical Physics and Center for NanoScience,}
\affiliation{${}^{\phantom{2}}$Ludwig-Maximilians-Universit\"{a}t M\"{u}nchen, 80333 M\"{u}nchen, Germany}
\affiliation{$^{4}$Institute for Theoretical
      Physics, University of Cologne, 50937 Cologne, Germany}
\affiliation{$^{5}$Institut N\'{e}el - CNRS and 
      Universit\'{e} Joseph Fourier, 38042 Grenoble, France}
\affiliation{$^{6}$Institut Universitaire de France, 
      103 boulevard Saint-Michel, 75005 Paris, France}
\affiliation{$^{7}$ Materials Science Division, 
      Argonne National  Laboratory, Argonne, Illinois 60439, USA}
\begin{abstract}
  We exploit the decoherence of
  electrons due to magnetic impurities, studied via weak
  localization, to resolve a longstanding question concerning
  the classic Kondo systems of Fe impurities in the noble metals
  gold and silver: which Kondo-type model yields a realistic
  description of the relevant multiple bands, spin and orbital
  degrees of freedom?  Previous studies suggest a fully screened
  spin $S$ Kondo model, but the value of $S$ remained ambiguous. We perform
  density functional theory calculations that suggest $S =
  3/2$. We also compare previous and new measurements of both the
  resistivity and decoherence rate in quasi 1-dimensional wires to
  numerical renormalization group predictions for $S=1/2,1$ and $3/2$, finding
  excellent agreement for $S=3/2$.
\end{abstract}
%\vspace{-1.0cm}
\date{\today}
\maketitle

{\em Introduction.---}
The Kondo effect of magnetic impurities in non-magnetic metals,
e.g. Mn, Fe or Co in Cu, Ag or Au, first manifested itself
in the early 1930's as an anomalous rise in resistivity with 
decreasing temperature, leading to a resistivity minimum 
\cite{deHaasvandenBergPhysica34}. In 1964 Kondo explained this effect
\cite{KondoProgTheorPhys64} as resulting from an antiferromagnetic 
exchange coupling between the spins of localized magnetic impurities 
and delocalized conduction electrons.
%
%shorten to get proofs down to 4 pages
%
%
%In the ensuing years the Kondo effect has become a paradigm for the interplay
%between localized and delocalized degrees of freedom and today is one
%of the best-studied problems in condensed matter physics with a wide range
%of applications. It is relevant, for example, for understanding 
%scanning tunneling microscopy spectra of magnetic Kondo adatoms on 
%surfaces\cite{Nagaoka02,Hirji07} or quantum transport through molecular 
%magnets\cite{Romeike06}. 

However, for many dilute magnetic alloys 
a fundamental question has remained unresolved to this day:
which effective low-energy Kondo-type model yields
a realistic description of the relevant multiple bands, spin and
orbital degrees of freedom \cite{NozieresBlandinJPhys80}?  
%
%
%Nevertheless, a fundamental aspect of the Kondo effect still
%remains poorly understood: for many dilute magnetic alloys it remains
%unclear to this day 
%which effective low-energy Kondo-type model yields
%a realistic description of the relevant multiple bands, spin and
%orbital degrees of freedom \cite{NozieresBlandinJPhys80}.  
Cases in
point are Fe impurities in Au and Ag, the former being the very first
magnetic alloy known to exhibit an anomalous resistivity minimum
\cite{deHaasvandenBergPhysica34}. 
Previous attempts to fit %the limited 
experimental data on, for
example, Fe impurities in Ag (%henceforth 
abbreviated as AgFe)
with exact theoretical results for thermodynamics,
by assuming a fully screened low-energy effective Kondo
model \cite{DesgrangesJPhysC85,HewsonBook97}, 
have been inconclusive: specific heat data is absent and
the local susceptibility of Fe in Ag obtained from
M\"ossbauer spectroscopy~\cite{SteinerHuefnerPRB75} indicated a spin
of $S=3/2$ while a fully screened $S=2$-model has been used to fit the
temperature dependence of the local
susceptibility~\cite{SchlottmannSacramentoAdvPhys93}. 
%The susceptibility
%curves for the different fully screened Kondo models are almost 
%indistinguishable \cite{DesgrangesJPhysC85}, making a determination 
%of the model from this quantity difficult.

A promising alternative route to identify
%towards identifying [MAKE THIS LATER IN CASE, PAPER STILL TOO LONG] 
the model for Fe in Au or Ag is offered by studying transport properties
of  high purity quasi-one dimensional mesoscopic wires of Au and Ag, doped with
a carefully controlled number of Fe impurities by means of
ion-implantation
\cite{PierreBirgePRB03,SchopferSaminadayarPRL03,BauerleSaminadayarPRL05,
MalletBauerlePRL06,AlzoubiBirgePRL06,CapronBauerlePRB08}.
Magnetic impurities affect these in two different ways. Besides causing the
afore-mentioned resistivity anomaly, they also make an anomalous
contribution $\gm (T)$ to the electronic phase decoherence rate
$\gammaphi (T)$ measured in weak (anti)localization: an itinerant
electron which spin-flip-scatters off a magnetic impurity leaves a
mark in the environment and thereby suffers decoherence.  By checking
model predictions for both effects against experimental observations
over several decades in temperature, decoherence can thus be harnessed
as a highly sensitive probe of the actual form of the effective
exchange coupling. Experiments along these lines
\cite{MalletBauerlePRL06,AlzoubiBirgePRL06} 
were consistent with a Kondo model in which the impurity spin is 
fully screened and inconsistent with underscreened or overscreened
Kondo models \cite{MalletBauerlePRL06}. A consistent description
of {\em both} resistivity and decoherence measurements using
the simplest fully screened Kondo model, the $S=1/2$ single-channel 
Kondo model, was, however, not possible: different Kondo scales were 
required for fitting the resistivity and decoherence rates 
\cite{MalletBauerlePRL06,AlzoubiBirgePRL06}.

In this Letter we address the above problem via the following 
%three-pronged 
strategy: (i) We carry out density functional theory
calculations within the local %(spin) 
density approximation (LDA) for
Fe in Au and Ag to obtain information that allows us to
prescribe a low-energy effective model featuring 3 bands coupling to
impurities with spin $S=3/2$. (ii) We calculate the resistivity $\RRm
(T)$ and decoherence rate $\gamma_{\rm m} (T)$ due to magnetic
impurities for three fully screened Kondo models, with $n=2S = 1$, 2 and 3,
using Wilson's numerical renormalization group (NRG) approach. (iii) We 
compare these predictions to 
%previous and new 
experimental data:
extracting the characteristic Kondo temperature $\TKS$ for each choice
of $n$ from  fits to $\RRm (T)$ and using these $\TKS$ to obtain
parameter-free predictions for $\gm (T)$, we find that the latter agree
best with experiment for $n=3$.

{\em LDA calculations.---} Fully relaxed density functional 
theory calculations %within LDA 
employing the VASP code
\cite{KresseFurthmuellerPRB96} showed that low-symmetry Fe
configurations (%such as 
split-interstitials \cite{VoglDederichsPRL76}) 
are energetically unfavorable:
%, implying that 
Fe impurities prefer an environment with cubic symmetry.
As the calculated defect formation energy of an Fe
interstitial was found to be about 2\, eV higher than the energy of a
substitutional defect, we discuss the latter case in the
following. This is 
%also 
in line with experiments on Fe-implantation in AgAu alloys, where
only substitutional Fe-atoms are found \cite{KirschFrotaPessoaEPL02}.

\begin{figure}[t]
\includegraphics[width=0.9\linewidth]{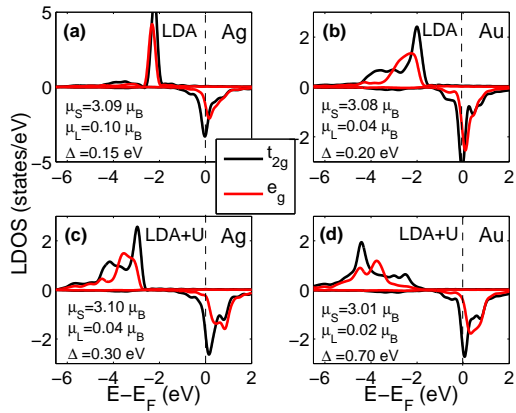}
\caption{ (color online).
  The d-level local density of states 
  (LDOS) of substitutional 
  Fe in Ag and Au within
    spin-polarized LDA (a,b) and LDA+U (c,d), with inclusion of spin-orbit
    interactions, and showing the
    e$_g$ (red) and t$_{\rm 2g}$ (black) components of the d-level 
    LDOS of FeAg (left panels) 
    and FeAu (right panels). Majority/minority contributions 
    are shown positive/negative. Legends give the spin
    ($\mu_{S}$) and orbital ($\mu_{L}$) magnetic moments in units of
    the Bohr magneton $\muB$, and the splitting ($\Delta$) between the
    e$_{\rm g}$ and t$_{\rm 2g}$ components of the d-level LDOS.
}
\label{fig1}
\end{figure}

Fig.~\ref{fig1} shows the d-level local density of states (LDOS) of substitutional Fe
in Ag and Au, obtained by spin-polarized calculations using a 108 atom
supercell, with similar results being found for a 256 atom supercell.
The cubic local symmetry
leads to e$_{\rm g}$ (doublet) and t$_{\rm 2g}$ (triplet) components
with a e$_{\rm g}$-t$_{\rm 2g}$ splitting, $\Delta\gtrsim 0.15$\,eV in
LDA (Fig.~\ref{fig1}(a-b)). The widths $\Gamma_{\rm e_g}$ and $\Gamma_{\rm t_{2g}}$ of
the e$_{\rm g}$ and t$_{\rm 2g}$ states close to the Fermi level
($E_{\rm F}$) are of the order of 1\,eV, resulting from a substantial
coupling to the conduction electrons.
The large t$_{\rm 2g}$ component at $E_{\rm F}$ persists within LDA+U 
(Fig.~\ref{fig1}(c-d) using $U=3$\,eV and a Hund's coupling $J_H=0.8$\,eV).

The spin and orbital moments are given in the legends of Fig.~\ref{fig1}
(spin-polarized Korringa-Kohn-Rostoker calculations
yielded similar values \cite{KirschFrotaPessoaEPL02}): Within
spin-polarized LDA a large spin moment $\mu_S$ of approximately
3-3.1\,$\muB$ forms spontaneously, consistent with M\"osbauer
measurements that give 3.1-3.2\,$\muB$ for the spin moment for Fe
in Ag \cite{SteinerHuefnerPRB75}.  In contrast, there is no tendency
for a sizable orbital moment (or a Jahn-Teller distortion). The small
orbital moments $\mu_L$ of $< 0.1 \, \muB$ (consistent with experimental
results \cite{brewer.04})  arise only due to weak spin-orbit coupling. 
We therefore conclude that the orbital degree of freedom is quenched on
an energy scale set by the width $\Gamma_{t_{\rm 2g}}$ of the t$_{\rm
  2g}$ orbitals. Moreover, since the spin-orbit splitting of the localized
spin in the cubic environment is proportional to $\mu_L^4$, it 
is tiny, well below our numerical precision of $0.01$\,meV, and, 
therefore, smaller than the relevant Kondo temperatures.

{\em Low-energy effective models.---}
The above results justify
formulating an effective low-energy model in terms of the spin-degree
of freedom only. The large spin moment $\mu_S$ of 3-3.1\,$\muB$ 
suggests an effective spin $S=3/2$.  Our LDA results thus
imply as effective model a spin-3/2 3-channel Kondo model, involving
local and band electrons of t$_{\rm 2g}$ symmetry.  An alternative
possibility, partially supported by the large (almost itinerant)
t$_{\rm 2g}$ component at $E_{\rm F}$, would be to model the system as
a spin-1 localized in the e$_{\rm g}$ orbitals, that is perfectly
screened by two conduction electron channels of e$_{\rm g}$
symmetry. This spin is then also coupled to (almost itinerant) t$_{\rm
  2g}$ degrees of freedom via the ferromagnetic $J_H$.  At high
temperature, the latter binds an itinerant $t_{\rm 2g}$ spin 1/2 to
the local spin 1 to yield an effective spin-3/2, consistent with the
spin-moment of 3-3.1 $\muB$ obtained within LDA, whereas in the low
temperature limit, the irrelevance of $J_{H}$ under renormalization
\cite{HewsonBook97} leads to the stated effective spin-1, 2-band
model. Though such a model is well justified only for $J_{H}\ll
\Gamma_{\rm t_{2g}}$, which is not the case here where $J_{H}\sim
\Gamma_{\rm t_{2g}}$, our LDA results do not completely exclude such a
model. To identify which of the models is most appropriate, we shall
confront their predictions with experimental data below.

%For the remainder of this Letter, we 
We thus describe 
Fe in Ag and Au using the following fully screened Kondo model:
\begin{equation}\label{eq:NChannelKondo}
   H = \sum_{k\alpha \alpha} \varepsilon_k c^\dagger_{k \alpha \sigma}
     c_{k \alpha \sigma} + J \sum_\alpha {\bf S} \cdot {\bf s}_\alpha \ .
\end{equation}
It describes $n$ channels of conduction electrons with wave vector
$k$, spin $\sigma$ and channel index $\alpha$, whose spin density
$\Sigma_\alpha {\bf s}_\alpha$ at the impurity site is coupled
antiferromagnetically to an Fe impurity with spin $S=n/2$. Whereas our
LDA results suggest $n=3$, we shall also consider the cases $n=1$ and 2.

%{\em Numerical renormalization group results:}
{\em NRG calculations.---}
The resistivity $\RRm (T)$ and decoherence rate $ \gm (T)$
induced by magnetic impurities can be obtained from the temperature 
and frequency dependence of the impurity spectral density 
\cite{MicklitzRoschPRL06,ZarandAndreiPRL04}. We have calculated 
these quantities using the NRG \cite{NRGspectra,
  NRGspectraDevelopments,WeichselbaumVonDelftPRL07}. 
While such calculations are routine for $n=1$ and 2 \cite{NRGspectra},
they are challenging for $n=3$. Exploiting recent advances in the NRG
\cite{WeichselbaumVonDelftPRL07} we were able to obtain accurate results
also for $n=3$ (using a discretization parameter of $\Lambda=2$ 
and retaining 4500 states per NRG iteration). 

\begin{figure}[t]
  \includegraphics[width=0.9\linewidth,angle=0]{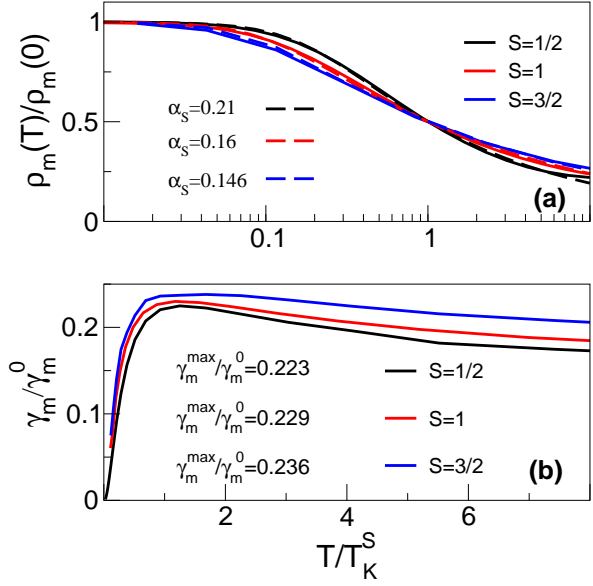}
\caption{ (color online).
(a) Resistivity $\RRm (T)$ (solid lines), and, 
  (b) decoherence rate $\gm (T)$ for $2S = n =1,2,3$; 
$\RRm(0)=2\tau\bar{\rho}/\pi\hbar\nu_{0}$,
$\gm^0=2/\pi\hbar\nu_{0}$, where $\bar{\rho}$ is the residual resistivity, 
$\nu_{0}$ the 
density of states
per spin and channel, $\tau$ the elastic scattering time, and 
$\gm^{\rm max}$ is the maximum value of $\gm (T)$.  We defined the
Kondo scale $\TKS$ for each $S$ via 
$\RRm(\TKS)=\RRm(0)/2$.  Dashed lines in (a) show that the
empirical form $\RRm(T)/\RRm(0)\approx f_{S}(T/\TKS)$ with
$ f_S(x)=(1+(2^{1/\alpha_S}-1)x^{2})^{-\alpha_S}$, used 
to fit experimental to NRG results for $S=1/2$ \cite{GoldhaberGordonMeiravPRL98}, 
also adequately fits the NRG results for $S=1$ and $S=3/2$.
}
\label{fig2}
\end{figure}

Fig.~\ref{fig2} shows $\RRm(T)$ and $\gm(T)$ for $n=2S=1,2$ and $3$. 
For $T \gtrsim \TKS$, enhanced spin-flip scattering causes 
both $\RRm (T)$ and $\gm(T)$ to increase with decreasing 
temperature.  For $T \lesssim \TKS$ the effective exchange 
coupling becomes so strong that the impurity spins are fully 
screened by conduction electrons, forming spin singlets, 
causing $\RRm (T)$ to saturate to a constant and $\gm(T)$ to 
drop to zero.  While these effects are well-known 
\cite{KondoProgTheorPhys64,PierreBirgePRB03,
SchopferSaminadayarPRL03,BauerleSaminadayarPRL05,
MalletBauerlePRL06,AlzoubiBirgePRL06}, it
is of central importance for the present study that they depend
quite significantly on $S = n/2$, in such a way that
\emph{conduction electrons are scattered and decohered more strongly
  the larger the local spin $S$}: With increasing $S$, (i) both
resistivities and decoherence rates decay more slowly with $T$ at
large temperatures $(\gg \TKS)$, and (ii) the ``plateau'' near
the maximum of $\gm (T)$ increases slightly in maximum height
$\gm^{\rm max}$ and significantly in width.  
These changes turn out to be sufficient to identify the proper
value of $S$ when comparing to experiments below.  

{\em Comparison with experiment.---}
We compared our theoretical results for
$\rho_{\rm m}(T)$ and $\gamma_{\rm m}(T)$ to measurements
on quasi 1-dimensional, disordered wires, for two AgFe
samples \cite{MalletBauerlePRL06}, (AgFe\,2 and AgFe\,3
having $27 \pm 3$ and $67.5 \pm 7$~ppm Fe impurities in Ag,
respectively), with a Kondo scale $\TK\approx 5{\rm K}$ 
(for $S=3/2$, see below).  These measurements extend up to
$T \lesssim \TK$ allowing the region $T/\TK \lesssim 1$
of the scaling curves in Fig.~\ref{fig2} to
be compared to experiment.  At $T\gtrsim \TK \approx 5$K
(i.e. $T/\TK\ge 1$) the large phonon contribution to the decoherence
rate prohibits reliable extraction of $\gamma_{\rm m}(T)$ 
for our Ag samples (see below) .  In order to compare theory and
experiment for temperatures $T/\TK\ge 1$, above the maximum in the
decoherence rate, we therefore carried out
new measurements on a sample (AuFe\,3) with $7 \pm 0.7$~ppm Fe impurities in
Au with a lower Kondo scale $\TK\approx 1.3${\rm K} but, as
discussed above, described by the same Kondo model.  Combining both
sets of measurement thereby allows a large part of the scaling
curves in Fig.~\ref{fig2} to be compared with experiment.

\begin{figure}
  \includegraphics[width=0.7\linewidth,angle=-90]{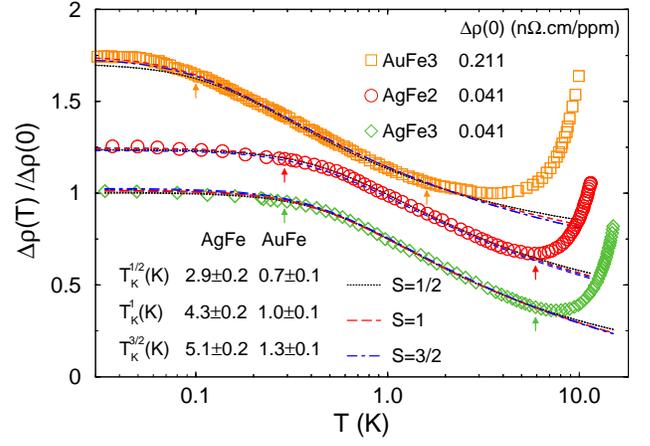}
\caption{ (color online).
Measured resistivities $\Delta\rho(T)$ (symbols) fitted to
equation (\ref{eq:fitting-ansatz-1}) (lines), for $n = 2 S=1, 2$ and
3, in the range below the onset of the
phonon contribution, but above $100-200$\,{\rm mK} \cite{note}. Specifically, 
we used $0.1-1.6$\,{\rm K} for AuFe and $0.29-5.9$\,{\rm
  K} for AgFe (arrows). The curves for AgFe\,2 and
AuFe\,3 have been offset vertically by 0.25 and 0.75, respectively.
The inset gives the Kondo scales $\TKS$ for AgFe and
AuFe extracted from the fits.  
Estimates of the 
unitary Kondo resistivities for $n=1, 2$ and $3$ (in units of
n$\Omega$.cm/ppm) yield $\rho_{\rm m} (0) = 0.041$, 0.047 and 0.049 for AgFe
(averaged over the two samples), and 0.23, 0.26 and 0.27 for AuFe,
respectively.
}
\label{fig3}
\end{figure}

Following \cite{MalletBauerlePRL06}, we subtract 
the electron-electron
contribution \cite{AltshulerKhmelnitskyJPhysC82} from the total 
resistivity $\rho$, yielding $\Delta \rho$ 
due to magnetic impurities (m) and phonons (ph):
\begin{equation}
  \label{eq:delta-rho}
  \Delta \rho(T) = \RRm(T) + \rho_{\rm ph} (T) + \delta \; .
\end{equation}
Here $\delta$ is an (unknown) offset \cite{note-unknown-offset}
and $\Delta \rho(T)$ is expressed per magnetic impurity. 
For temperatures low enough that $\rho_{\rm ph}(T)$
can be neglected, $\Delta\rho(T)-\delta$ 
%is the quantity of interest, corresponding 
corresponds to the theoretical curve
$\rho_{\rm m} (T) = \rho_{\rm m} (0) f_S (T/\TKS)$ [cf.\ caption of
Fig.~\ref{fig2}], where $\rho_{\rm m} (0) =\Delta\rho(0)-\delta $
is the unitary Kondo resistivity.  Fig.~\ref{fig3} illustrates how
we extract the Kondo scale $\TKS$ and $\rho_{\rm m} (0)$ from the 
experimental data, by fitting the
Kondo-dominated part of $\Delta \rho (T)$ in a fixed temperature range
(specified in the caption of Fig.~\ref{fig3}) to the NRG results of
Fig.~\ref{fig2}(a), using the Ansatz
\begin{equation}
  \label{eq:fitting-ansatz-1}
\Delta \rho(T) \approx \delta + 
(\Delta\rho(0) - \delta)f_{S}(T/\TKS) \; . 
\end{equation}
Such fits are made for each of the fully screened Kondo models, 
using $\TKS$ and $\delta$ as fit parameters. Importantly,
the values for $\TKS$ and $\rho_\subi(0)$ obtained from the fits,
given in the inset and caption of Fig.~\ref{fig3}, respectively, show a
significant $S$-dependence: both $\TKS$ and $\rho_\subi (0)$
increase with $S$, since the slope of the logarithmic Kondo increase
of the theory curves for $\rho_{\rm m}$ [cf.\ Fig.~\ref{fig2}] decreases
significantly in magnitude with $S$. Nevertheless, all three models
fit the Kondo contribution very well, as shown in
Fig.~\ref{fig3}, so a determination of the appropriate model from resistivity 
data alone is not possible.

\begin{figure}[t]
  \includegraphics[width=0.7\linewidth,angle=-90]{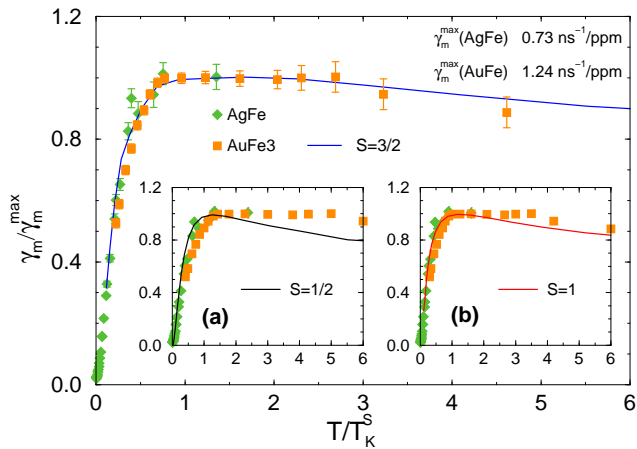}
\caption{ (color online).
    Comparison of the measured (symbols) and theoretical (lines)
    results for the dimensionless decoherence rate $\gm (T)
    / \gm^{\rm max}$ as function of $T/T^S_{K}$, using $S=3/2$.
    Insets show comparisons to $S=1/2$ (a) and $S=1$ (b). 
    $\TKS$ for AgFe and AuFe was extracted from the resistivities
    (inset of Fig.~\ref{fig3}), while $\gamma_{\rm
      m}^{\rm max}$ was determined as the average plateau height in
    the region $T/\TK^{3/2} \in [0.7,1.35]$.  
    Typical error bars are shown for $S=3/2$. They grow with
    increasing temperatures due to the increasing difficulty of
    subtracting the growing phonon contribution to the decoherence
    rate.
}
\label{fig4}
\end{figure}

To break this impasse, we exploit the remarkably sensistive
$S$-dependence of the spin-flip-induced decoherence rate
$\gamma_{m}(T)$. Fig.~\ref{fig4} shows the
measured  dimensionless decoherence rate $\gm
(T) / \gm^{\rm max}$ for Ag and Au samples (symbols) as function of
$T/\TKS$ for $S=1/2, 1 $ and $3/2$, using the $\TKS$ values extracted
from the resistivities, together with the corresponding
\emph{parameter-free} theoretical predictions (lines), taken from
Fig.~\ref{fig2}(b).  The agreement between theory and experiment is poor for
$S=1/2$, better for $S=1$ but excellent for $S=3/2$, confirming the
conclusion drawn above from {\em ab initio} calculations.  The
dependence on $S$ is most strikingly revealed through the width of the
plateau region (in units of $T/\TKS$), which grows with $S$ for the
theory curves but shrinks with $S$ for the experimental data (for
which $\TKS$ grows with $S$), with $S=3/2$ giving the best agreement.

%
%New shortened conclusions
%
%
{\em Conclusions.---} 
In this Letter we addressed one of the fundamental
unresolved questions of Kondo physics: that of deriving and
solving the effective low-energy Kondo model appropriate for a realistic
description of Fe impurities in Au and Ag. Remarkably, for both Ag and Au samples, 
the use of a fully screened $S=3/2$ three channel Kondo model allows a 
\emph{quantitatively consistent} 
description of both the resistivity and decoherence rate {\em with a
single $\TK$} (for each material).
Our results set a benchmark for the level of quantitative understanding 
attainable for the Kondo effect in real materials.
 
%The remarkable fact that for both Ag and Au samples, the use of an
%$n = 2S = 3$ fully screened Kondo model allows a \emph{quantitatively consistent} 
%description of both the resistivity and decoherence rate {\em with a
%single $\TK$} (for each material), extracted from the resistivities,
%is the central result of this work. Our results set a benchmark for 
%the level of quantitative understanding attainable for the Kondo effect 
%in real materials.
%
%For shortened version drop the following as it repeats what we added
%in the introduction
%
%It contrasts to previous work
%\cite{MalletBauerlePRL06,AlzoubiBirgePRL06}, where the use of a
%$S=1/2$ Kondo model could fit either the resistivity or the
%decoherence rate using a given $\TK$, but not both.

%
%old conclusion, before shortening to PRL 4 page limit
%
%{\em Conclusions:}
%In this Letter we addressed one of the fundamental
%unresolved questions of Kondo physics: that of deriving and
%solving the effective low-energy Kondo model appropriate for a realistic
%description of Fe impurities in Au and Ag. Guided by information 
%from new {\em ab initio} calculations we proposed a fully 
%screened, 3-band spin-3/2 Kondo model as answer and solved this
%for dynamical quantities, for the first time, by exploiting recent 
%advances in the NRG method. To substantiate our suggestion, we undertook a
%systematic theoretical study of three fully screened Kondo models
%with $n=2S = 1, 2$ and 3. The results allowed us to make 
%reliable predictions for the dependence on $S$ of the
%resistivity and, crucially, also of the Kondo-induced decoherence
%rate. This combined $S$-dependence from both quantities was
%sufficiently pronounced to determine $S=3/2$ as the
%appropriate choice, through comparison with previous and new
%experimental data. 

L. B. acknowledges support from the EU within the
Marie Curie Actions for Human Resources and Mobility; P.M. 
from the ESF %under the 
programme SONS, 
contract N. ERAS-CT-2003-980409; T. M. from the 
U.S. Dept. of Energy, Office of Science, 
%under 
Contract No. DE- AC02-06CH11357;
L.S. and C.B. acknowledge technical support from the Quantronics group, 
Saclay and A.D. Wieck %from Bochum University 
and financial 
support from ANR PNANO "QuSPIN". Support from 
the John von Neumann 
Institute for Computing (J\"ulich), the DFG (SFB 608, SFB-TR12 and De730/3-2) and
from the Cluster of Excellence \emph{Nanosystems Initiative Munich} is 
gratefully acknowledged.
%\vspace{-0.7cm}

\end{document}